\documentstyle[12pt,psfig]{article}
\begin{document}
\begin{titlepage}
\begin{center}

{\Large\bf{The Color Evaporation Model in Diffractive $J/\psi$ Photoproduction}}
\\[5.0ex]
{\Large\it{ M. B. Gay  Ducati $^{*}$\footnotetext{$^{*}$E-mail:gay@if.ufrgs.br}}}\\
 {\it and}\\
{ \Large \it{ C. B. Mariotto $^{**}$\footnotetext{$^{**}$E-mail:mariotto@if.ufrgs.br} 
}} \\[1.5ex]
{\it Instituto de F\'{\i}sica, Univ. Federal do Rio Grande do Sul}\\
{\it Caixa Postal 15051, 91501-970 Porto Alegre, RS, BRAZIL}\\[5.0ex]
\end{center}

{\large \bf Abstract:}
Our goal is the study of color effects on the $J/\psi$ and $D$ mesons production, as well as the study of their elastic production.We use the Color Evaporation Model (CEM), where a single factor takes into account the probability that the $c\overline{c}$ produced in a color octet state turns into a physical (colorless) quarkonium or continues as a color octet, producing open charm.

The main result of this work, besides investigating the range of applicability of the CEM, is to show that this model can be used to describe the elastic processes of $J/\psi$ and open charm photoproduction without the use of a Pomeron.

\vspace{1.5cm}



\end{titlepage}

\section{Introduction}

The study of the heavy quarks production is still a challenge, both theoretically and experimentally. The masses of the heavy quarks give additional scales, where the perturbative QCD is applicable. This allows one to test the QCD, that is the theory candidate to explain the strong interactions. Data on mesons with heavy quarks (specially with charm) have been accumulated, which make possible the support of the theory on experimental results.

Besides, the production of exclusive charmed mesons involves the interplay between short and long distance physics. The question of where the color effects should be included, whether in the perturbative or in the non-perturbative part, is mandatory and for now model dependent. An analysis based on a comparison among different theoretical predictions surveying the experimental results was made in \cite{dissertacao}. In fact, the Color Singlet Model (CSM) \cite{KramerNLO} fails to explain the quarkonium production at the TEVATRON \cite{braatenfleming}, hadroproduction experiments \cite{BenekeRothstein} and the production via $Z^0$ decay \cite{ZdecaysOct}, whereas both the Color Octet Model (COM) \cite{braatenfleming, BenekeRothstein, ZdecaysOct} and the Color Evaporation Model (CEM) \cite{Halzen, HALZENquantit, ZdecaysGreg} have successful predictions in all theses processes. In inelastic $J/\psi$ photoproduction, the COM can reasonably describe the HERA data \cite{heradata} if higher order QCD effects due to the multiple emission of gluons \cite{Y}, or effects of intrinsic $k_T$ smearing \cite{X} are considered.
The CEM also gives a reasonable description of these data once one imposes a cutoff $|\hat{t}|>4m_c^2$ ($m_c$ is the charm mass) on the invariant mass of the virtual gluon \cite{CEMinelast}. These facts motivate us to do further investigations about the color effects on charmonium production, which in this work are done in the framework of the CEM.
 
Another question very discussed currently is how the elastic and diffractive scattering takes place. These events are characterized by the presence of rapidity gaps - regions on the rapidity spectrum where no hadron is detected, and correspond to $10\%$ of the events at HERA \cite{10porcento}. There are two main approaches for treating this problem: in the first one, namely the diffractive models, the object exchanged in the interaction must be color singlet and have the vacuum quantum numbers - the so-called Pomeron; in the second one, namely the soft color interactions, the Pomeron is unnecessary. The Color Evaporation Model is a soft color model, which was developed to explain the quarkonium production. However, it has been discussed its application in elastic and diffractive events. In this case, not only do the soft color interactions eliminate the color of the $c\overline{c}$ pair, but also characterize the elastic process at large distances.

The basic questions that we try to answer are the following: which is the mechanism for elastic production? Are those mechanisms alternative?
We outline the answer in the case of charmed mesons photoproduction.
The main point is to show that the CEM can be used to describe elastic photoproduction of $J/\psi$ and $D$ mesons, extending the limit of its applicability.

This work is divided as follows: We start with the description of the Color Evaporation Model (section \ref{CEMlabel}). In section \ref{secModDifrat}, we present a diffractive model applied to elastic $J/\psi$ photoproduction. In section \ref{minjoben}, we present our results, applying the CEM to elastic $J/\psi$ and open charm photoproduction. We also test the CEM in a LO hadroproduction calculation. In section \ref{conc} we present our conclusions.

\section{The Color Evaporation Model}
\label{CEMlabel}

In the Color Evaporation Model \cite{Halzen, HALZENquantit}, the production of charmonium states involves two stages: a $c\overline{c}$ pair is produced (perturbative process) and then evolves non-perturbatively into the asymptotic states. 

The diagrams of the hard part of the process to be considered are taken into account irrespective of the $c\overline{c}$ color and spin-parity. Color-octet states can also contribute, and its color (as well as its spin-parity) is neutralized through soft gluons exchanges in the large-distance part of the process. Since the soft gluons carry very low momentum, the dynamics is not affected. The color effects that characterize the octet to singlet transition are therefore considered in the non-perturbative part of the process, and the information about the quantum numbers of the perturbative $c\overline{c}$ is lost. 

The Color Evaporation Model predicts that the cross section for hidden and open charm production is proportional to the rate of production of a $c\overline{c}$ pair integrated in an appropriate mass range  \cite{Halzen},
\begin{eqnarray}
\sigma_{onium}&=& \frac{1}{9}\int_{2m_c}^{2m_D}dM_{c\overline{c}}
	\frac{d\sigma_{c\overline{c}}}{dM_{c\overline{c}}} 
\label{1} \\
\sigma_{open}&=& \frac{8}{9}\int_{2m_c}^{2m_D}
dM_{c\overline{c}}
	\frac{d\sigma_{c\overline{c}}}{dM_{c\overline{c}}}
	+ \int_{2m_D}^{\sqrt s}dM_{c\overline{c}}
	\frac{d\sigma_{c\overline{c}}}
{dM_{c\overline{c}}} \,\,,
\label{2}
\end{eqnarray}
where $d\sigma_{c\overline{c}}/dM_{c\overline{c}}$ is calculable perturbatively, $M_{c\overline{c}}$ is the invariant mass of the $c\overline{c}$ pair, $m_c$ is the charm quark mass, and $2m_D$ is the $D\overline{D}$ threshold. The factors $\frac{1}{9}$ and $\frac{8}{9}$ represent the statistical probabilities that the $3\otimes\overline{3}=1\oplus8$ $c\overline{c}$ pair will be in a singlet or octet state asymptotically. If singlet, it contributes to the charmonium production; if octet, it binds to a light quark, thus producing open charm. In addition, if the $c\overline{c}$ invariant mass is above the $D$ threshold, its energy is too large for the binding to occur, and therefore also contributes to the open charm production.

Each specific state of charmonium carries one fraction of the overall charmonium produced,
\begin{eqnarray}
\sigma_i=\rho_i\,\sigma_{onium}\,\,\,\,\,\,  
(i=J/\psi, \chi, \psi^{\prime},...) \,.
\label{3}
\end{eqnarray} 
The fraction $\rho_i$ is a non-perturbative input, and then it has to be determined by fitting the data. Since it describes fractions of the overall charmonium produced, it should be independent of the perturbative process that originated the charmonium. This universality makes the CEM to have a predictive power. Once $\rho_i$ is determined, it could be used to predict other processes with no additional parameters.

\section{The Diffractive Models}
\label{secModDifrat}

Elastic processes are currently studied in terms of diffractive models \cite{Ryskin}. In these models, one assumes that the interaction between the photon and the nucleon takes place through the exchange of a Pomeron - a color singlet object with the vacuum quantum numbers. With this assumption the photon and the $J/\psi$ quantum numbers are identical and the elastic scattering is characterized. Besides, the rapidity gaps are explained by this mechanism. In the model of Ryskin {\it et al.} \cite{Ryskin}, the Pomeron is treated perturbatively as composed of two gluons in the leading logarithm approximation or of a gluons ladder in higher order, which is specially important in the small $x$ region \cite{Ryskin}.

In the proton rest frame the cross section factorizes in three parts corresponding to the following sequence of events: fluctuation of the photon into a $c\overline{c}$ pair, interaction between this pair and the proton, and subsequent binding of the $c\overline{c}$ into the $J/\psi$. In the small $x$ region, this factorization follows from the fact that the $\gamma \rightarrow c\overline{c}$ and $c\overline{c} \rightarrow J/\psi$ processes last a much longer time than the intermediary interaction process. The $\gamma \rightarrow c\overline{c}$ and the interaction amplitudes can be calculated in perturbation theory, whereas the $c\overline{c} \rightarrow J/\psi$ process is non-perturbative and expressed by the $J/\psi$ wave function $\psi^J(c\overline{c})$.

The Pomeron exchange makes the cross section to be proportional to the squared gluon distribution and therefore sensitive to the gluon parameterization. For this reason it has been argued that elastic $J/\psi$ photoproduction is a good place to probe the gluon content in the proton \cite{Ryskin}. 

The results then obtained for the cross section of the process explain its behavior but not the normalization. This is a consequence of the uncertainty in the $J/\psi$ wave function, necessary in this approach, which makes the results somewhat unreliable.

\section{The Color Evaporation Model in Elastic Processes}
\label{minjoben}

The soft color models \cite{Buchmuller, SCI, eboli9708283} are an alternative approach to the Pomeron based models. The basic idea is that the object exchanged in the $t$ channel can be an ordinary QCD object such as a (colored) gluon. The parton combinations that emerge from the hard interaction in a color octet state can participate in soft color interactions with the remaining partons of the proton. As a consequence, singlet objects are produced in the final state and rapidity gaps are observed.

These models are based on the fact that soft color interactions can exchange color. Since spin is no longer conserved by the $c\overline{c}$ system in the transition into the quarkonium state, this assumption can violate the heavy quark spin symmetry, that is a principle believed for a long time \cite{IsgurWise}. However, no rigorous QCD proof of this symmetry was given up to now.

Our proposal is, based on the phenomenological success of the CEM, to describe the elastic processes of $J/\psi$ and open charm photoproduction in the framework of the Color Evaporation Model.

\subsection{The Elastic Photoproduction of $J/ \psi$}

In section \ref{secModDifrat} we have seen briefly the usual description of elastic $J/\psi$ photoproduction through a diffractive model \cite{Ryskin}. It is predicted that the cross section increases proportionally to the squared gluon distribution. In contrast, in the CEM the exchanged object is of the same nature the one in events without rapidity gaps (it can be a color octet one), which results in a cross section linearly proportional to the gluon distribution. Therefore, it seems premature to extract the gluon distribution at this level before confirming which model is the correct one.

In a preceding calculation \cite{HALZENquantit}, Amundson {\it et al.} have used the CEM in photo and hadroproduction of $J/\psi$, in a NLO calculation, up to ${\cal O}(\alpha\alpha_s^2)$ and ${\cal O}(\alpha_s^3)$, respectively ($\alpha=\frac{1}{137}$ and $\alpha_s$ is the QCD coupling constant). With a specific set of parameters - $\rho_{J/\psi}=0.5\,\,(0.43)$, $m_c=1.45\,GeV\,\,(1.43\,GeV)$, factorization scale $\mu_F=m_c\,(2m_c)$ and GRV gluon distribution \cite{GRV} (MRS-A \cite{MRS}) in the proton, they obtained a good description of the data \cite{dadosgreg} in both processes. Then they concluded that the CEM gives an accurate prediction once they have done a calculation in NLO.

Our purpose is to calculate the elastic process of $J/\psi$ photoproduction without using a Pomeron. In this case, the soft interactions have a double function: to eliminate the color of the $c\overline{c}$ pair and to make the elastic scattering viable.

The main contribution comes from the subprocess $\gamma g\rightarrow c\overline{c}$. In the case of the LO contribution, all energy of the photon is transferred to the $J/\psi$ - there is no hard gluon emitted, which would carry part of the momentum of the photon and would form a jet, making the process an inelastic one. The result of the perturbative calculation is well known for a long time \cite{Gluck78}. All we have to do is to consider the soft interactions effect and to restrict the region of the invariant mass of the $c\overline{c}$ that contributes to the bound states of the physical $c\overline{c}$ (charmonium). These ideas are the basis of the calculation that follows.

With the assumptions above, it turns out that the cross section to the elastic $J/\psi$ photoproduction is given by (\ref{1}), where the differential cross section is
\begin{eqnarray}
\frac{d\sigma_{\gamma g\rightarrow c\overline{c}}}
{dM_{c\overline{c}}} = \frac {4\pi\alpha\alpha_se_c^2}
{M^2_{c\overline{c}}} [(1+\gamma+\frac{1}{2}\gamma^2)\ln(\frac{1+\sqrt{1-\gamma}}
{1+\sqrt{1-\gamma}}) \nonumber \\
-(1+\gamma) \sqrt{1-\gamma}]\,\,g_P(x, \mu)\,\,, 
\label{integrandoELAST} 
\end{eqnarray}
where $e_c$ is the charm charge, $\gamma=\frac{4m_c^2}{M_{c\overline{c}}^2}$, $M^2_{c\overline{c}}={\hat{s}}=xs$, $s$ is the squared CM energy of the $\gamma p$ system, $g_P(x, \mu)$ is the gluon distribution inside the proton, and $\mu$ is the factorization scale. According to the CEM, only contributes to the charmonium production the region below the open charm threshold, $2m_D$ (and trivially the region above the minimum invariant mass to produce a $c\overline{c}$ pair). In the region $M_{c\overline{c}}>2m_D$, the energy of the system is too large to allow the $c$ and the $\overline{c}$ quarks to remain in a bound state.

The statistic effect of the soft interactions is accounted by the $1/9$ factor, which represents the probability that the $c\overline{c}$ pair is in a color singlet state. That is the other condition for the $c\overline{c}$ pair to remain bound.

Finally, the $\rho_{J/\psi}$ factor gives the fraction of the total charmonium that contributes to the $J/\psi$. In this factor both the direct production and the indirect contributions (from decays of higher mass charmonium states - $\psi^{\prime}$, $\chi$, $\eta_c(2S)$) are included. That is the free parameter of the model.

Our application of the CEM is, however, free of parameters - we used the same set of parameters as the previously determined by Amundson {\it et al.} \cite{HALZENquantit}, namely $m_c=1.45\,GeV$, $\rho_{J/\psi}=0.5$, $\mu_F=(M_{c\overline{c}})^{1/2}$, and the GRV gluon distribution \cite{GRV}. In doing so, we are testing the universality of the model, expressed by the non-perturbative factor $\rho_{J/\psi}/9$. 

Our results are shown in fig. \ref{opencharmmej}. We obtained an excellent agreement with the data of elastic processes \cite{frabetti, H1, ZEUS}, including the normalization of the cross section. It is worth to stress that although the behavior of the HERA elastic data \cite{H1, ZEUS} (to which our prediction best agrees) is fitted by the Ryskin diffractive model, uncertainties in the $J/\psi$ wavefunction do not allow an absolute prediction in normalization. On the other hand, once the CEM does not require a $J/\psi$ wavefunction, this uncertainty is not present in our calculation.

Thus, we got two basic results: the universality of the parameter $\rho_{J/\psi}/9$ was confirmed for another process, namely elastic $J/\psi$ photoproduction, and this elastic process can be described without the use of the Pomeron.

\subsection{The Elastic Photoproduction of $D$ Mesons}

As we have seen in section $2$, the CEM describes the charmonium and the open charm production by the same dynamics, that is, by the same perturbative processes. Thus, the question of whether the elastic process can be described without a Pomeron is also applied in this case. In the positive case, we will have two evidences that the Pomeron is unnecessary in elastic processes.

By supposing the same processes as in the case of charmonium, we have to consider only the contributions of the regions of invariant mass complementary to the charmonium one. As a result the cross section for the production of open charm can be found from the expressions (\ref{2}) and (\ref{integrandoELAST}).
In this case, two regions contribute:

- If the $c\overline{c}$ pair is an octet state, which is expressed by the $8/9$ factor, the $c\overline{c}$ pair does not remain bound, and the $c$ and the $\overline{c}$ bind to a light (anti)quark forming $D$ mesons separately. 

- In the region $M_{c\overline{c}}>2m_D$, the pair is pushed apart, because its energy is too large for the binding to occur. Therefore, $D$ mesons are formed in this region even though the $c\overline{c}$ pair is a singlet state asymptotically. This is expressed in the second integral of  (\ref{2}), where no probabilistic factor is present.

Our results are shown in figure \ref{opencharmmej}, where we discriminated the contribution coming from the region $2m_c<M_{c\overline{c}}<2m_D$, and its sum with the superior region, giving the total cross section. We obtained that both terms of (\ref{2}) are important and competitive, contrary to the assumption in the literature \cite{Halzen}. The results are compared with the data of elastic processes \cite{dadosgreg}. With no free parameter, agreement is noticeable both in behavior and in normalization. This gives another evidence that the CEM can explain elastic processes, currently treated through the Pomeron exchange.

As an additional test of the CEM, we repeat the procedure of Amundson {\it et al.}, by employing the CEM to describe the $J/\psi$ and open charm hadroproduction. By simplicity, we only use the LO calculation instead of the NLO. The processes that contribute in LO are the gluon fusion $gg \rightarrow c\overline{c}$ and light quark-antiquark annihilation, $q\overline{q} \rightarrow c\overline{c}$. The differential cross section is given by
\begin{eqnarray}
\frac{d\sigma_{pp \rightarrow c\overline{c}}}{dM_{c\overline{c}}}(M_{c\overline{c}})
= \int_{0}^{1-M^2_{c\overline{c}}/s} dx_F
\frac{1}{s(x_F^2+4M^2_{c\overline{c}}/s)^{1/2}} \nonumber\\
(\sum_q {\sigma}_{q\overline{q} \rightarrow Q\overline{Q}}(M_{c\overline{c}})
(q^A (x_A (x_F, M_{c\overline{c}}), M_{c\overline{c}}){\overline{q}}^B (x_B (x_F, M_{c\overline{c}}), M_{c\overline{c}})
+ q \leftrightarrow \overline{q}) \nonumber \\
+\sigma_{gg \rightarrow Q\overline{Q}}(M_{c\overline{c}})
 g^A (x_{A} (x_{F}, M_{c\overline{c}}), M_{c\overline{c}}) g^B (x_{B} (x_{F}, M_{c\overline{c}}), M_{c\overline{c}})) \,\,.
\label{integrandoHADRO} 
\end{eqnarray}
where $M^2_{c\overline{c}}=x_Ax_Bs$ is the squared invariant mass of the $c\overline{c}$ pair, $x_F=x_A-x_B$, $q^{A(B)}$ and $g^{A(B)}$ are the quarks and gluons distributions in the hadron $A(B)$, ${\sigma}_{q\overline{q} \rightarrow Q\overline{Q}}$ and $\sigma_{gg \rightarrow Q\overline{Q}}$ are the respective partonic cross sections \cite{Gluck78}.

Our results are shown in figure \ref{psihadromej}, where we plot the cross sections for $J/\psi$ (full curve) and D mesons (dotted line) versus data \cite{dadosgreg}. Again, the two contributions on the different regions of invariant mass (dotted-dashed and dashed lines) are competitive, their sum resulting in the total cross section. By using the same parameters $\rho_{J/\psi}=0.5$ and $m_c=1.45\,GeV$ as in photoproduction, we describe the $D$ mesons data with a factor $K=8$. This same factor is then used for the $J/\psi$.

Our results show that the proportionality of the $J/\psi$ and $D$ mesons cross sections was maintained, and the predictions give a dependence with the CM energy in agreement with the data. However, the absolute normalization was not far enough to explain the data, which means that the NLO corrections can be large. In fact, even the NLO calculation requires a $K$ factor of about $1.5$ \cite{HALZENquantit}. This may indicate that the hadroproduction processes have additional effects like radiation emitted by the initial state hadrons, which remain to be investigated more accurately.

\section{Conclusion}
\label{conc}

In this letter the mechanisms of elastic production of charmed vector mesons are studied. We discussed the diffractive model of Ryskin {\it et al.} in this context, where a Pomeron is considered in lowest order as two gluons and in higher order as a gluon ladder. We have also described a mechanism of soft color, namely the Color Evaporation Model, where non-perturbative interactions characterize the elastic scattering at large distances. These two models predict a different dependence of the cross section on the gluon distribution.

Motivated by the phenomenological successes of the CEM, we propose its application in the elastic processes of $J/\psi$ and open charm photoproduction, once considered the convenient diagrams without the emission of hard gluons. The most important point is that due to the soft interactions there is no need to introduce the Pomeron exchange. 

Our results are in accordance with the data of elastic processes of charmonium and open charm photoproduction, showing that the soft interactions can successfully replace the Pomeron in these cases. This suggests further investigation whether the CEM can describe other elastic processes currently described by the Pomeron, like jet production for instance. 

\section{Acknowledgements}

The authors acknowledge useful comments of G. Ingelman. This work was partially financed by CNPq and by Programa de Apoio a N\'ucleos de Excel\^encia (PRONEX), BRAZIL.

\newpage

\vspace{1.0cm}

\section*{Figure Captions}

\vspace{1.0cm}

\begin{figure}[h]
\psfig{file=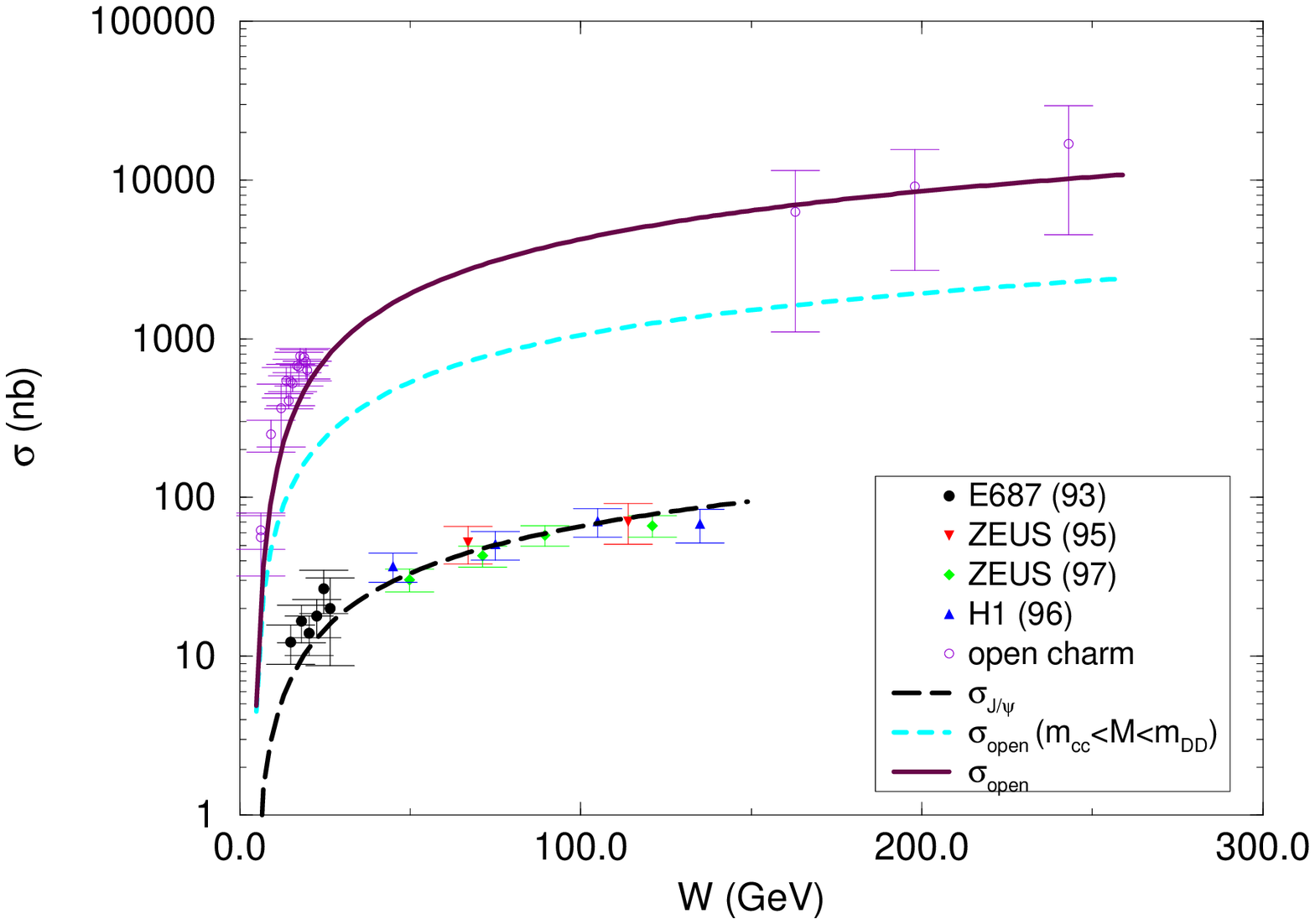,width=130mm}
\caption{}
\label{opencharmmej}
\end{figure}

Fig. \ref{opencharmmej}: Results of the Color Evaporation Model for the cross section versus CM energy for $J/\psi$ (dashed line) and open charm photoproduction (full line) in comparison with the elastic data \cite{frabetti, H1, ZEUS, dadosgreg}. In the open charm case, the contribution for the same invariant mass range as the charmonium case is discriminated. The predictions are absolute in normalization.

\vspace{1.0cm}

\begin{figure}[h]
\psfig{file=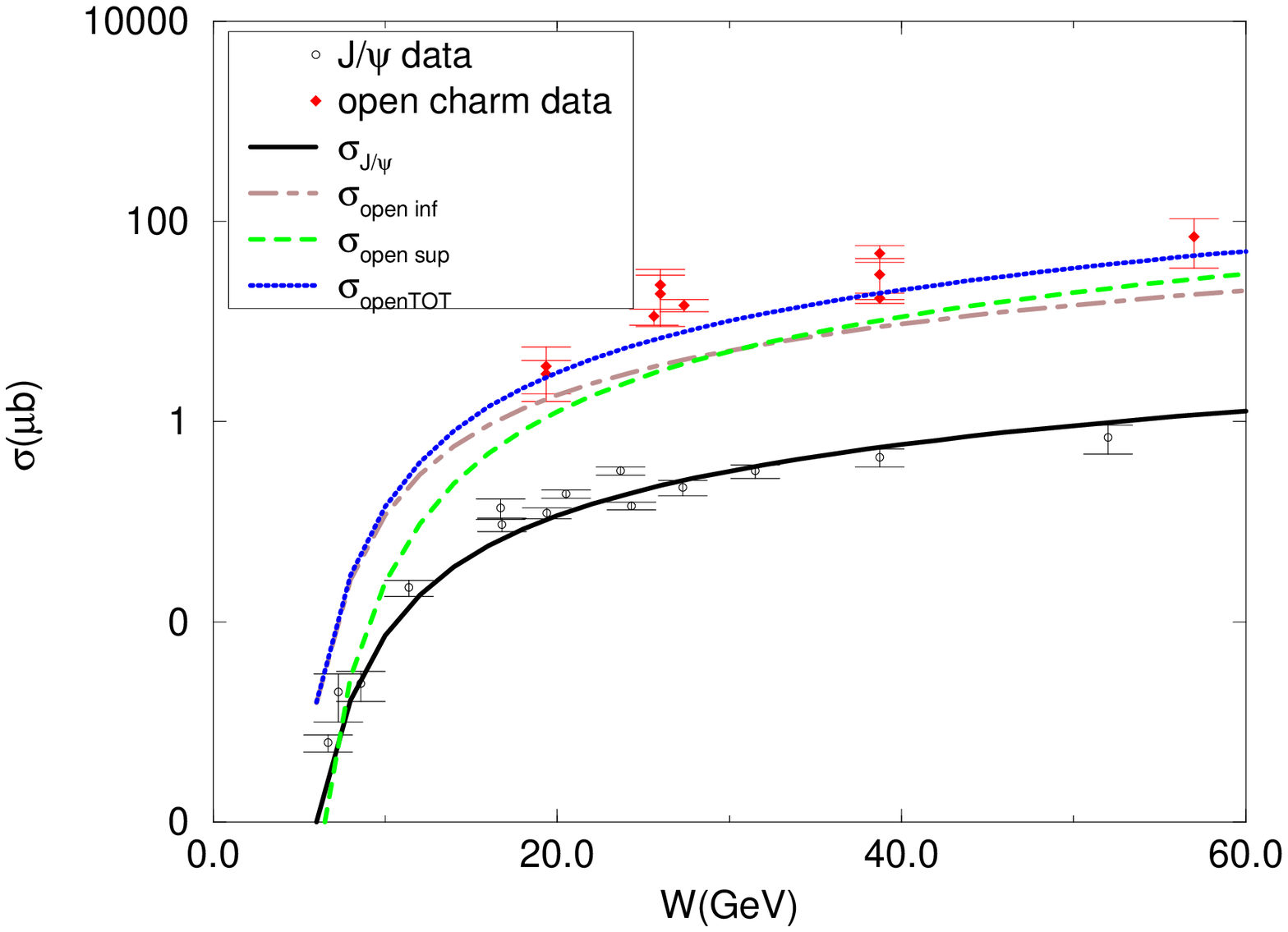,width=130mm}
\caption{}
\label{psihadromej}
\end{figure}

Fig. \ref{psihadromej}: The results of the LO hadroproduction calculation for $J/\psi$ and open charm versus CM energy, in comparison with experimental results \cite{dadosgreg}. Using the same parameters as in the photoproduction case, a factor $K=8$ which takes into account the higher order corrections is required.

\newpage

\newpage 

\begin{thebibliography}{99}

\bibitem{dissertacao} Mariotto, C. B., {\it Color and Mass Effects on Heavy Quark Production}, Master Sciences Dissertation, IF-UFRGS, Porto Alegre (1999) (in portuguese).

\bibitem{KramerNLO} Kr\"amer, M.,  {\it Nucl. Phys.} {\bf B459} (1996) 3; Berger, E. L., Jones, D., {\it Phys. Rev.} {\bf D23} (1981) 1521.


\bibitem{braatenfleming}
Braaten, E., Fleming, S., {\it Phys. Rev. Lett.} {\bf 74} (1995) 3327.


\bibitem{BenekeRothstein}
Beneke, M., Rothstein, I. Z., {\it Phys.Rev.} {\bf D54} (1996) 2005; Erratum-ibid. {\bf D54} (1996) 7082.

\bibitem{ZdecaysOct}
Ernstr\"om, P., L\"onnblad, L., V\"anttinen, M., {\it Z. Phys.} {\bf C76} (1997) 515. 

\bibitem{Halzen}
Amundson, J. F., \'Eboli, O. J. P., Gregores, E. M., Halzen, F., {\it Phys. Lett.} {\bf B 372} (1996) 127.

\bibitem{HALZENquantit}
Amundson, J. F., \'Eboli, O. J. P., Gregores, E. M., Halzen, F., {\it Phys. Lett.} {\bf B 390} (1997) 323.

\bibitem{ZdecaysGreg} 
Gregores, E. M., Halzen, F., \'Eboli, O. J. P., {\it Phys. Lett.} {\bf B 395} (1997) 113.

\bibitem{heradata}
H1 Collab., Aid, S. {\it et al.}, {\it Nucl. Phys.} {\bf B472} (1996) 3;  ZEUS Collab., Breitweg, J. {\it et al.}, {\it Z. Phys.} {\bf C76} (1997) 599.

\bibitem{Y}
Kniehl, B. A., Kramer, G., {\it Eur. Phys. J} {\bf C6} (1999) 493; Kniehl, B. A., hep-ph/9907315.

\bibitem{X}
Sridhar, K., Martin, A. D., Stirling, W. J., {\it Phys. Lett.} {\bf B 438} (1998) 211.


\bibitem{CEMinelast}
\'Eboli, O. J. P., Gregores, E. M., Halzen, F.,  {\it Phys. Lett.} {\bf B 451} (1999) 241.


\bibitem{10porcento}
H1 Collab., Ahmed, T. {\it et al.}, {\it Nucl. Phys.} {\bf B435} (1995) 3;
H1 Collab., Derrick, M. {\it et al.}, {\it Phys. Lett.} {\bf B346} (1995) 399.


\bibitem{Ryskin} Ryskin, M. G., Roberts R. G., Martin, A. D., Levin, E. M., {\it Z. Phys.} {\bf C76} (1997) 231. 


\bibitem{Buchmuller} Buchm\"uller, W., Hebecker, A., {\it Phys. Lett.} {\bf B355} (1995) 573.

\bibitem{SCI} Edin, A., Ingelman, G., Rathsman, J., {\it Phys. Lett. B} {\bf 366} (1996) 371.

\bibitem{eboli9708283} \'Eboli, O. J. P., Gregores, E. M., Halzen, F., {\it Nucl. Phys. Proc. Suppl.} {\bf 71} (1999) 349. 

\bibitem{IsgurWise}
Isgur, N., Wise, M. B., {\it Phys. Lett.} {\bf B 232} (1989) 113.

\bibitem{GRV} Gl\"{u}ck, M., Reya, E., Vogt, A., {\it Z. Phys.} {\bf C67} (1995) 433.

\bibitem{MRS} Martin, A. D., Roberts, R. G., Stirling, W. J., {\it Phys. Lett. B} {\bf 354} (1995) 155. 

\bibitem{dadosgreg} See \cite{HALZENquantit} and references therein.

\bibitem{Gluck78}
Gl\"{u}ck, M., Reya, E., {\it Phys. Lett.} {\bf 79} (1978) 453.

\bibitem{frabetti} 
Frabetti, P. L. {\it et al.}, {\it Phys. Lett.} {\bf B 316} (1993) 197.

\bibitem{H1}
H1 Collab., Aid, S. {\it et al.}, {\it Nucl.Phys.} {\bf B472} (1996) 3.

\bibitem{ZEUS} 
ZEUS Collab., Derrick, M. {\it et al.}, {\it Phys. Lett.} {\bf B 350} (1995) 120; ZEUS Collab., Breitweg, J. {\it et al.}, {\it DESY 97-060} (1997).

\end{thebibliography}
\end{document}